\documentclass[11pt]{article} % Default font size is 12pt, it can be changed here

\usepackage{geometry} % Required to change the page size to A4

\usepackage{graphicx} % Required for including pictures

\usepackage{float} % Allows putting an [H] in \begin{figure} to specify the exact location of the figure
\usepackage{wrapfig} % Allows in-line images such as the example fish picture

\usepackage{lipsum} % Used for inserting dummy 'Lorem ipsum' text into the template

\usepackage{amsmath}
\usepackage{xcolor}

\usepackage{lmodern}
\usepackage[T1]{fontenc}
\usepackage{cfr-lm}

\usepackage{sectsty}
\sectionfont{\rmfamily\mdseries\itshape\Large}
\subsectionfont{\rmfamily\mdseries\itshape\large}
\subsubsectionfont{\rmfamily\mdseries\itshape}

\usepackage{setspace} 
\usepackage[labelfont=bf,labelsep=period,justification=raggedright]{caption}

\usepackage{soul}

\makeatletter
\renewcommand{\@biblabel}[1]{\quad#1.}
\makeatother

\date{}

\begin{document}

\newcommand{\HRule}{\rule{\linewidth}{0.25mm}} % Defines a new command for the horizontal lines, change thickness here

% Title must be 150 characters or less
\begin{center}
\LARGE{Trees of Unusual Size: Biased Inference of Early Bursts from Large Molecular Phylogenies}\\

% Insert Author names, affiliations and corresponding author email.

\Large{Matthew W. Pennell$^{1,2,3,4}$, Brice A.J. Sarver$^{1,2,3}$ \& Luke J. Harmon$^{1,2,3}$}\\[0.2cm]

\end{center}

\begin{center}
$^{1}$Institute for Bioinformatics and Evolutionary Studies (IBEST) and $^{2}$Department of Biological Sciences, University of Idaho
\\
$^{3}$BEACON Center for the Study of Evolution in Action
\\
$^{4}\mathtt{mwpennell@gmail.com}$\\[0.5cm]
\HRule\\[2cm]
\end{center}

% Please keep the abstract between 250 and 300 words
\section*{Abstract}

An early burst of speciation followed by a subsequent slowdown in the rate of diversification is commonly inferred from molecular phylogenies. This pattern is consistent with some verbal theory of ecological opportunity and adaptive radiations. One often-overlooked source of bias in these studies is that of sampling at the level of whole clades, as researchers tend to choose large, speciose clades to study. In this paper, we investigate the performance of common methods across the distribution of clade sizes that can be generated by a constant-rate birth-death process. Clades which are larger than expected for a given constant-rate branching process tend to show a pattern of an early burst even when both speciation and extinction rates are constant through time. All methods evaluated were susceptible to detecting this false signature when extinction was low. Under moderate extinction, both the $\gamma$-statistic and diversity-dependent models did not detect such a slowdown but only because the signature of a slowdown was masked by subsequent extinction. Some models which estimate time-varying speciation rates are able to detect early bursts under higher extinction rates, but are extremely prone to sampling bias. We suggest that examining clades in isolation may result in spurious inferences that rates of diversification have changed through time.

% Please keep the Author Summary between 150 and 200 words
% Use first person. PLoS ONE authors please skip this step. 
% Author Summary not valid for PLoS ONE submissions.   
%\section*{Author Summary}

\section*{Introduction}
The branching patterns of reconstructed molecular phylogenies contain information about the tempo and mode of evolution\cite{Raup1973}, \cite{Nee1992}. This insight has been invaluable to our understanding of many evolutionary processes and patterns. Recent studies have identified a common pattern of apparent slowdowns in the rate of diversification through time (reviewed in \cite{GavriletsLosos2009}, \cite{Glor2010}). Such a pattern is consistent with some theoretical work on adaptive radiations, based on the idea that diversification rates will decrease as species fill available niches \cite{Simpson1953}, \cite{Schluter2000}, \cite{Yoder2010} (but see \cite{McPeek2008} for an alternate interpretation). Phylogenies with slowdowns have been inferred for diverse groups of organisms (e.g. lizards \cite{Harmon2003}, birds \cite{Weir2006}, \cite{PhillimorePrice2008}, fish \cite{Ruber2005}, and many groups of plants \cite{Davies2004}). 

The most commonly used metric for detecting shifts in the rate of diversification is the $\gamma$-statistic introduced by Pybus and Harvey \cite{PybusHarvey2000}. This statistic quantifies how internode distances (i.e. waiting times to speciation) vary through time compared to what one would expect under a pure-birth model of diversification. Under the pure-birth null expectation, $\gamma$ is distributed according to a standard normal distribution. A $\gamma$-value less than -1.645 (one-tailed test) represents a statistically significant slowdown in diversification rate. Positive values of $\gamma$ can be caused by either speed-ups in diversification rates or species turnover as recently diverged lineages have have not been around long enough to have been ``pruned'' by extinction, creating an overabundance of nodes closer to the present (``pull of the present"; \cite{Nee1992}). These two scenarios cannot be distinguished with the $\gamma$-statistic, so most studies follow the authors' \cite{PybusHarvey2000} original recommendation to disregard significantly positive $\gamma$-values. 

There has been considerable controversy surrounding the interpretations of slowdowns using the $\gamma$-statistic on molecular phylogenies. A preponderance of studies have inferred significantly negative $\gamma$- values across a wide range of taxonomic groups. McPeek \cite{McPeek2008} collected a large number of phylogenies from the literature and found that 80\% of clades had $\gamma < 0$ and 42\% had $\gamma < -1.96$. However, a number of other studies have pointed out that negative $\gamma$-values can result from factors other than slowdowns in diversification rates. It has been shown that $\gamma$ can be biased by under-parameterization of the model of sequence evolution \cite{Revell2005} and non-random taxon sampling \cite{Cusimano2010} (see \cite{Brock2011} and \cite{Cusimano2012} for potential solutions to this problem). Recent speciation events will have a similar effect; if speciation is modelled as a process rather than as a singular event, this can lead to apparent slowdowns in diversification rates \cite{EtienneRosindell2012}, \cite{Rosenblum2012}.

A recent simulation study, conducted by Liow et al. \cite{Liow2010}, found that $\gamma$ tended to be positively biased when trees were simulated under a birth-death process with variable rates of speciation. They found that, even when present, slowdowns in net diversification rates were very difficult to detect using $\gamma$, as short branches near the tips of the tree tend to obscure the signal of the slowdown. Under the conditions simulated by Liow et al. \cite{Liow2010}, it should be only possible to observe a slowdown for a small subset of values of speciation rate ($\lambda$) and extinction rate ($\mu$) and if lineages are sampled at a particular time in the clade's history. It seems unrealistic to propose that these two conditions are met for the majority of phylogenies. Nevertheless slowdowns are commonly inferred from empirical data. The reasons for the discordance between the analysis of empirical data and expectations derived from simulation studies are poorly understood and warrant further investigation.   

Several new modeling approaches have been developed to detect non-homogeneous diversification. One is to use a model in which speciation and/or extinction vary through time \cite{Rabosky2006Evolution}, \cite{Rabosky2006Evolution}, \cite{Morlon2010}, \cite{Stadler2011}, \cite{Morlon2011}. Diversity-dependence (in which speciation rate varies as a function of the number of taxa at a given time) has also been proposed as an explanation for the patterns of species richness \cite{Rabosky2009}. This has recently been modeled in a number of studies (e.g. \cite{RaboskyLovette2008PRSB}, \cite{Etienne2012}) to look for signatures of an adaptive radiation. Diversity-dependence is a useful approach as it is seemingly consistent with theory on adaptive radiation, which posits that diversification should slow as niches become filled \cite{Schluter2000}, \cite{Yoder2010}. There is also some evidence from the fossil record \cite{Alroy2008} (but see \cite{Benton2007} for an opposing perspective) to support this modeling approach. However, such models from both paleobiology and comparative biology are susceptible to the criticism that the ecological mechanisms by which diversity dependence would operate on the scale of a clade are not entirely clear \cite{Wiens2011}. While the mathematics of the various methods differ, they all involve fitting non-homogeneous birth-death models to phylogenies rather than using summary statistics such as the $\gamma$-statistic. The behavior and performance of these methods have not been explored in as much detail as the $\gamma$-statistic.

One important consideration that is often overlooked in studies of clade diversification is that our inferences may be biased by the way we choose clades to investigate. There are several types of sampling bias that are likely to be important. Cusimano and Renner \cite{Cusimano2010}, Brock et al.\cite{Brock2011} and Cusimano et al.\cite{Cusimano2012} investigated systematists' tendency to sample representative taxa, leading to an overabundance of nodes deep in the tree. However, we should also consider that we are only sampling clades that survive to the present day. This means that the observable distribution of surviving trees is only a subset of all possible trees \cite{Slowinski1989},\cite{Slowinski1993}. Another form of bias is that researchers interested in the adaptive radiations are likely to be interested in the speciose clades. These clades likely reside in the `tails' of the distribution of possible clade sizes and thus give a biased sample for the inference of diversification rates \cite{PhillimorePrice2008}. This effect of sampling clades has seldom been addressed in the literature (but see \cite{Slowinski1989},\cite{Slowinski1993}).  Inferences of early bursts are particularly problematic because even under a constant pure-birth or birth-death process, large clades are likely to show patterns of a rapid diversification early in their history. This is simply due to the fact that in order to be large, they are more likely to have undergone a stochastically high rate of speciation early in the process and subsequently regressed to the mean speciation rate \cite{Price2007}, \cite{PhillimorePrice2008}. Analyzing large clades in isolation may lead to inferring ecological processes (such as adaptive radiations) attributable solely to these stochastic processes. 
  
Phillimore and Price \cite{PhillimorePrice2008} investigated the influence of clade size and age on the $\gamma$-statistic though their simulations were limited to a small set of parameter space. Specifically, they conditioned their simulations on tree age to investigate the correlation between $\gamma$ and clade size and on clade size to investigate the correlation between $\gamma$ and tree age. As a result, the simulations tended to produce trees where the distribution of the parameter of interest was centered on the expected value. Here, we are specifically interested in investigating the statistical properties of trees that are \textit{unexpected}, that is, unusually large. By simultaneously conditioning our phylogenies on both tree age and size, we were able to obtain trees from the `tails' of the distribution. In addition to using the $\gamma$-statistic, we investigate the effects of the non-random sampling of clades using a model selection approach. We show that the effect of this bias can be severe but it affects different methods in different ways, and we make recommendations on the appropriate use of these methods.
% Results and Discussion can be combined.

\section*{Results and Discussion}
For all phylogenies simulated, regardless of the value of extinction rate used, large phylogenies are disproportionally likely to have undergone an initial burst of speciation events \cite{Stadler2008}, \cite{PhillimorePrice2008}. This can be visualized with a lineages-through-time plot (Figure 1). This result has consequences for the inference of diversification rate patterns; as the clades we choose to examine are not a random sample of clades but often tend to be interesting and speciose and, at least for molecular phylogenies, have survived to the present day. It is also the case that these large clades are the ones most amenable to statistical analysis.

\begin{figure}[!ht]
\begin{center}
\includegraphics[width=6in]{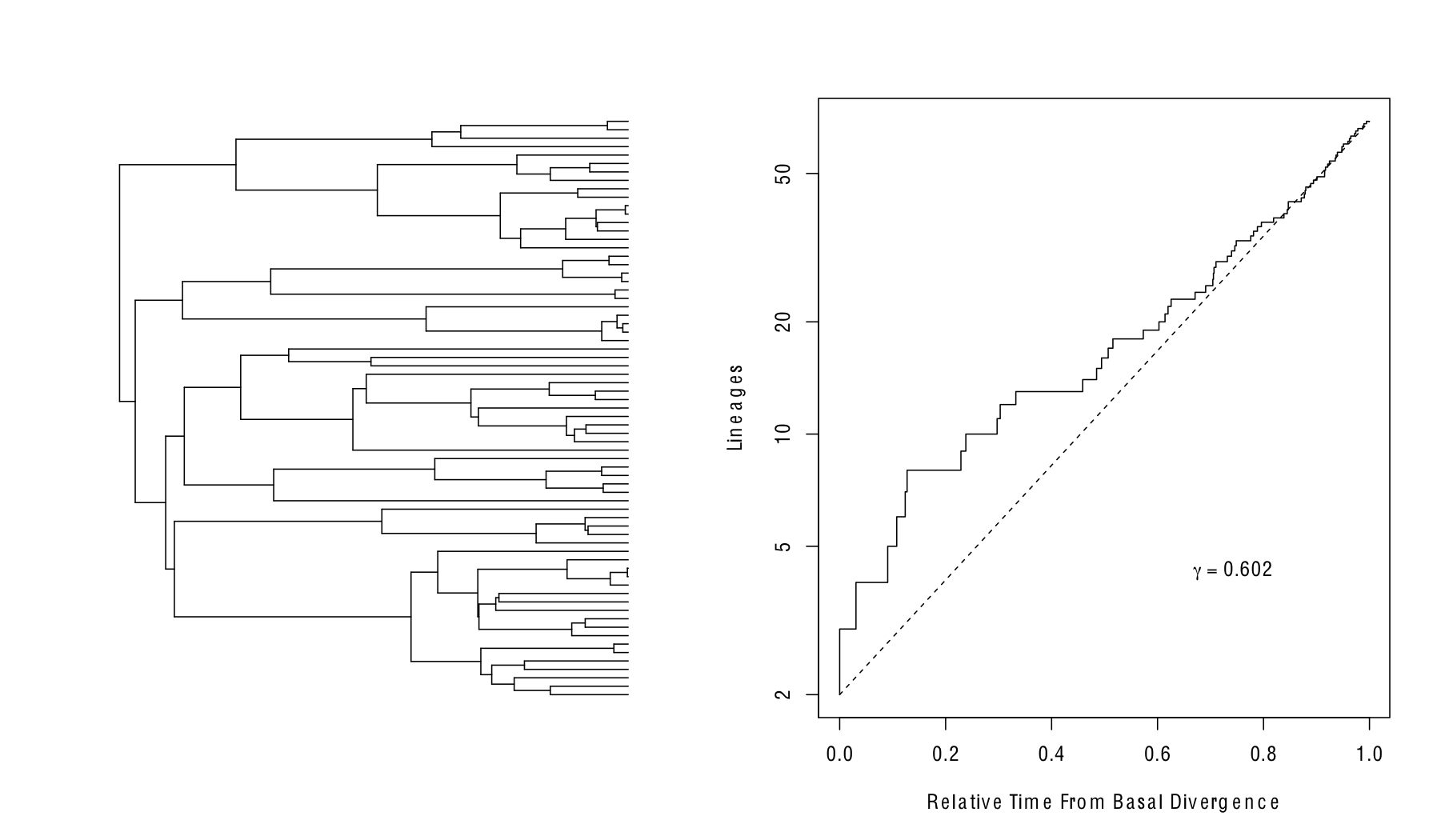}
\end{center}
\begin{flushleft}
\caption{
\textbf{Figure 1. Exemplar Lineages-Through-Time Plot.} An example of a lineages-through-time (LTT) plot for a tree (shown on left) drawn from the far right `tail' of the distribution of tree sizes (5 percent of surviving trees are expected to be this large or larger) for $\lambda=1$, $\mu=0.5$ and $\tau=5$. The dotted line is the expected number of lineages under a constant diversification rate. This LTT plot shows the typical signature of an `early burst' of speciation yet this signature is not captured by the $\gamma$-statistic ($\gamma = 0.602$; not significant) as the burst is `masked' by later extinction events.} 
\end{flushleft}
\end{figure}

We simulated phylogenies under constant-rate diversification, conditioning on both the number of taxa in the clade ($N$) and the age of the tree ($\tau$). We did this in order to investigate the statistical properties of trees drawn from different parts of the distribution of possible tree sizes that can result from a constant-rate process. Consistent with previous work \cite{Stadler2008}, \cite{PhillimorePrice2008}, trees which are large at the present day (i.e. those drawn from the `right tail' of the distribution) are more likely to harbor a signal of a slowdown using the $\gamma$-statistic even when the phylogeny is generated under a constant-rate process with low extinction. This is evident in Figure 2, where data points corresponding to tree sizes larger than the expectation (denoted by a dashed line) show elevated Type-1 error rates when simulated under low extinction rates. However this is not the case when background extinction rates are higher as extinction alters the distribution of branch lengths on the reconstructed phylogeny. The resulting distribution effectively obscures any signal of a burst (in our simulations, the burst is entirely attributable to stochastic processes) that might have occured deep in the tree \cite{Weir2006}, \cite{RaboskyLovette2008Evolution}, \cite{Liow2010}. This is especially true for large phylogenies (Figure 2); $\gamma$ is a summary statistic for all nodes in the tree and therefore the few early branching events will have an even smaller influence on the calculation of the $\gamma$-statistic for large trees with proportionally more nodes.
 
\begin{figure}[!ht]
\begin{center}
\includegraphics[width=6in]{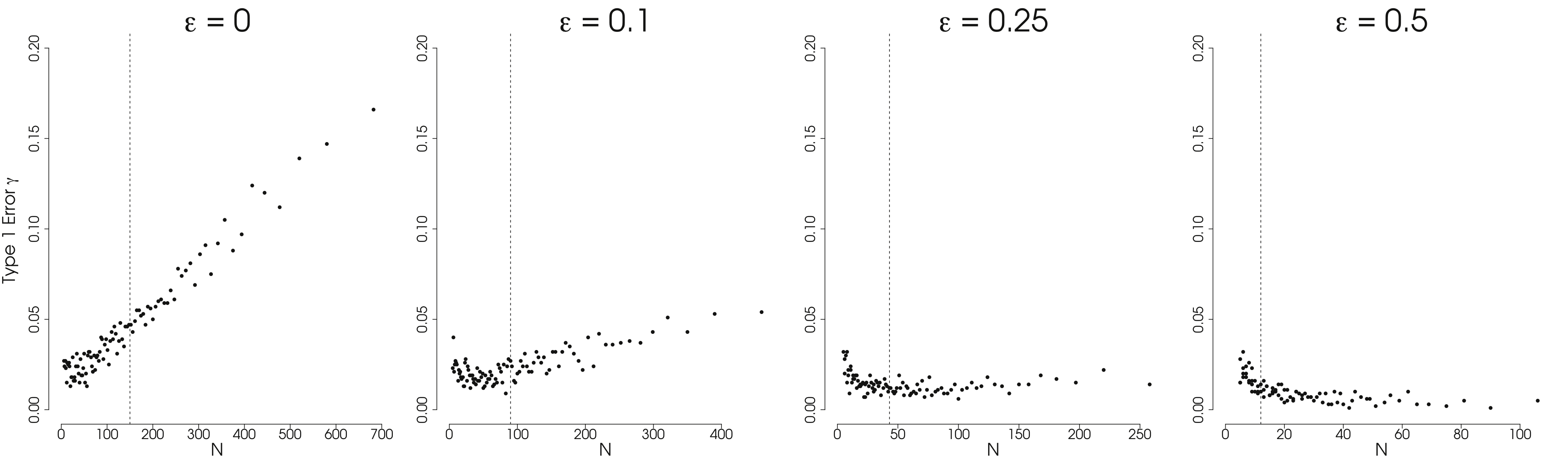}
\end{center}
\begin{flushleft}
\caption{
\textbf{Figure 2. Type-1 Error Rate for the $\gamma$-Statistic.} Results from simulations showing the Type-1 error rate for the $\gamma$ statisitc, signifying a false inference of a slowdown. All trees were generated under a constant-rate birth-death (or pure-birth) process. We recognize a Type-1 error if the value of $\gamma < -1.645$ (significant at $p=0.05$; one-tailed test). Extinction rate, $\epsilon=\mu/\lambda$, varies across the plots ($\epsilon=0,0.1,0.25,0.5$); speciation rate, $\lambda$, and total tree-depth, $\tau$ are held constant ($\lambda=1$ and $\tau=5$). All are plotted against the expected number of taxa across the cumulative distribution of probability densities (from 0.99 to 0.01). The dashed vertical line represents the expected value for $N$ under the simulating conditions. Each point represents 1000 simulations. (Results for $\epsilon =0.75$ and $\epsilon =1$ not shown.)}  
\end{flushleft}
\end{figure} 
  
Our results suggest two things about the use of the $\gamma$-statistic. First, the test has very low power to detect changes in speciation rates when species turnover rates are even modestly high \cite{Liow2010}. Second, while there is some concern that significantly negative $\gamma$-values may potentially be misleading \cite{Revell2005}, \cite{Cusimano2010}, \cite{Fordyce2010}, bias in sampling large clades does not tend to create false signatures of a slowdown under modest extinction rates. We know from the fossil record that extinction rates have been moderately high for most groups. This implies that, even if slowdowns in diversification rates are relatively common, significantly negative $\gamma$-values should be rare. However, this is not the case \cite{McPeek2008}. The widely reported evidence for slowdowns in studies using $\gamma$ is thus somewhat paradoxical and deserves explanation. Pessimistically, we may attribute this to some yet unstudied artifact such as sampling effort. More intensive sampling of lineages would have the effect of increasing the number of nodes close to the present. This would deteriorate any signal of an early burst detectable by the $\gamma$-statistic. It should be noted however, that a number of the phylogenies that provide evidence for a slowdown are near complete at the species level (e.g. \cite{PhillimorePrice2008}), though there may still be additional lineages that should actually be recognized as species. It may be the case that if more lineages were included, many of the signatures of early bursts in the literature would no longer be present. This is difficult to evaluate except on a case-by-case basis.

We found that the diversity-dependent models are prone to bias due to sampling in a similar manner as the $\gamma$-statistic (Figure 3); when death rates are low, the diversity-dependent model was preferred to a constant birth-death model more frequently for clades at the large end of the distribution of clade size. But even with moderate extinction, the diversity-dependent model was rarely preferred. In fact, for higher extinction rates, the reverse was true--larger clades were less likely to fit a diversity-dependent model than smaller clades (Figure 3). However, it should be noted that the model of diversity-dependence we used here did not explicity model extinction. Consequently the model estimates diversification rate as a function of contemporaneous lineages in the reconstructed phylogeny rather than as a function of contemporaneous lineages in the true (unobserved) phylogeny \cite{Bokma2009}. Recently Etienne et al. \cite{Etienne2012} provided a full likelihood solution for diversity-dependence with extinction which uses a Hidden-Markov approach to fit the model to a phylogeny. We justify the use of a diversity-dependent model which does not estimate extinction in our simulation study due to the fact that it is currently the more commonly employed approach. The diversity-dependent model thus has the same drawback as the $\gamma$-statistic. Extinction will tend to erode the signal of the early diversification events, especially so for large trees. The results from the time-varying speciation rate model (Figures 4 and 5; discussed below) suggest that this result will not hold if both extinction rates and speciation rates are explicitly modeled. However, it should be noted that there is some evidence that estimates of extinction can be inconsistent when there is rate variation across clades \cite{Rabosky2010}.

\begin{figure}[!ht]
\begin{center}
\includegraphics[width=6in]{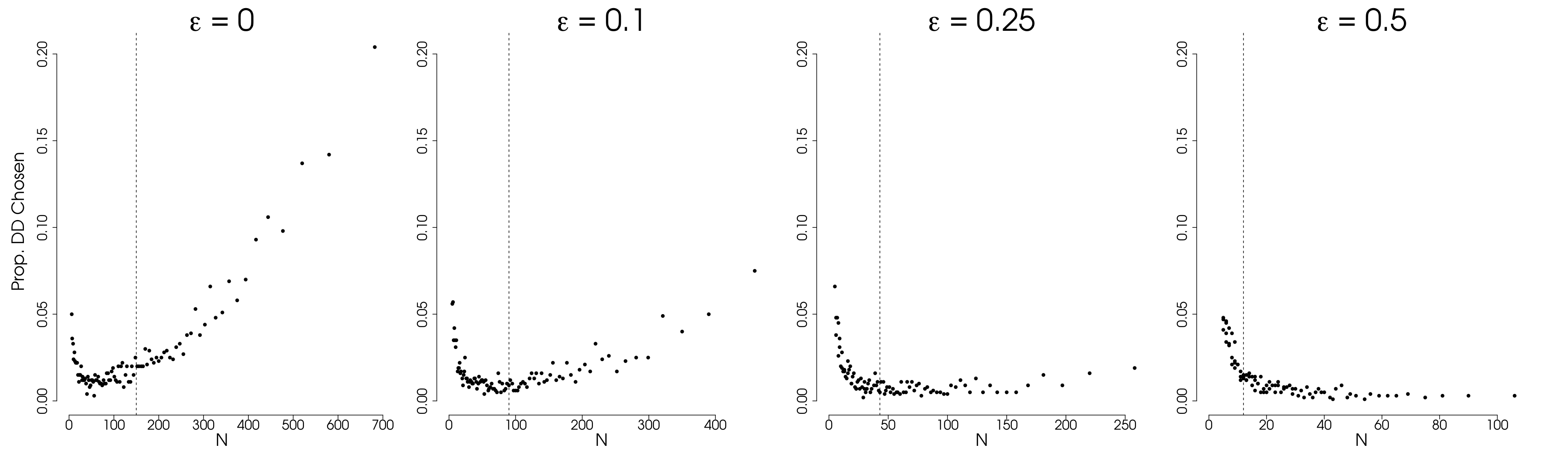}
\end{center}
\begin{flushleft}
\caption{
\textbf{Figure 3. Proportion of Trees Showing Support for Diversity-Dependent Model.} Results from simulations showing the proportion of phylogenies for which a density-dependent (DD) model is preferred over a constant-rate birth-death (BD) model in using AIC to select amongst the models. Only the DD model and a BD model were compared. We used a $\Delta$AIC cutoff of 4 to favor a DD model when the generating model was a constant-rate process. Extinction rate, $\epsilon=\mu/\lambda$, varies across the plots ($\epsilon=0,0.1,0.25,0.5$); speciation rate, $\lambda$, and total tree-depth, $\tau$ are held constant ($\lambda=1$ and $\tau=5$). All are plotted against the expected number of taxa across the cumulative distribution of probability densities (from 0.99 to 0.01). The dashed vertical line represents the expected value for $N$ under the simulating conditions. Each point represents 1000 simulations. (Results for $\epsilon =0.75$ and $\epsilon =1$ not shown.)}
\end{flushleft}
\end{figure}  

A correction that has been employed to increase the power to detect slowdowns for both the $\gamma$-statistic and diversity-dependent models is to collapse some nodes close to the present. Species delimitation is a problem that has recently received a great deal of attention from molecular systematists (e.g. \cite{OMeara2010},\cite{YangRannala2010},\cite{Kubatko2011}). Subdividing lineages into subspecies will necessarily increase the effect of the pull of the present and further obscure evidence of a slowdown. Collapsing these recent nodes will certainly influence our ability to infer slowdowns and this procedure has been done by some researchers (e.g. \cite{PhillimorePrice2008}, \cite{Slater2010}), by regarding recently-diverged species as not being ``good" species. However, there is currently a lack of theory to guide such decision-making. We therefore recommend that authors should be cautious in collapsing nodes as we do not fully understand how species delimitation affects these methods.

The time-varying speciation models \cite{RaboskyLovette2008Evolution}, \cite{Morlon2010} were preferred for data generated under a constant-rate process with increasing frequency as tree size increased. We found this to be true even when extinction was present (Figures 4 and 5), though the effect was dampened as extinction increased. As stated above, large trees generated under a constant rate birth-death process are subject to having undergone stochastic bursts of speciation in order to obtain their current size. The time-varying speciation rate models are sensitive to these bursts, which is both a positive and negative; positive because they have more power to detect changes in diversification through time and negative because these models are more prone to inferring spurious results as a consequence of sampling bias. We took two alternative approaches to compare model fits of a constant-rate birth-death model versus a time-varying speciation model: the full-likelihood approach of Rabosky and Lovette \cite{RaboskyLovette2008Evolution} and an approximate approach based on the coalescent process, recently derived by Morlon et al. \cite{Morlon2010} (see Methods for details). Contrary to our expectations, we found that these two approaches differed substantially in terms of which model was prefered when using Akaike Information Criterion (AIC). For trees from the larger end of the distribution, the coalescent approach tended to provide support for a time-varying speciation model over a constant-rate model at much higher frequencies than the full likelihood approach, especially when extinction rates were low (Figure 5). The reasons for these differences are not entirely clear. While many of the large trees do show `early bursts' due to stochastically high rates early in the process (see Figure 1 for an example), the proneness of the coalescent-based approach to favor a model more complex than the generating model is worrisome as sampling bias appears to very strongly influence model choice. Both approaches are relatively new and their respective statistical properties have not been explored at all beyond the publications in which they were presented \cite{RaboskyLovette2008Evolution}, \cite{Morlon2010}. The statistical properties of these and other related models is something that certainly warrants further investigation. Researchers are increasingly fitting more complex models of diversification to study diversity dynamics through time and across clades (e.g. \cite{Alfaro2009}, \cite{Morlon2011}, \cite{Stadler2011}) and it is imperative that we have a good understanding of these if we are to make meaningful inferences.

\begin{figure}[!ht]
\begin{center}
\includegraphics[width=6in]{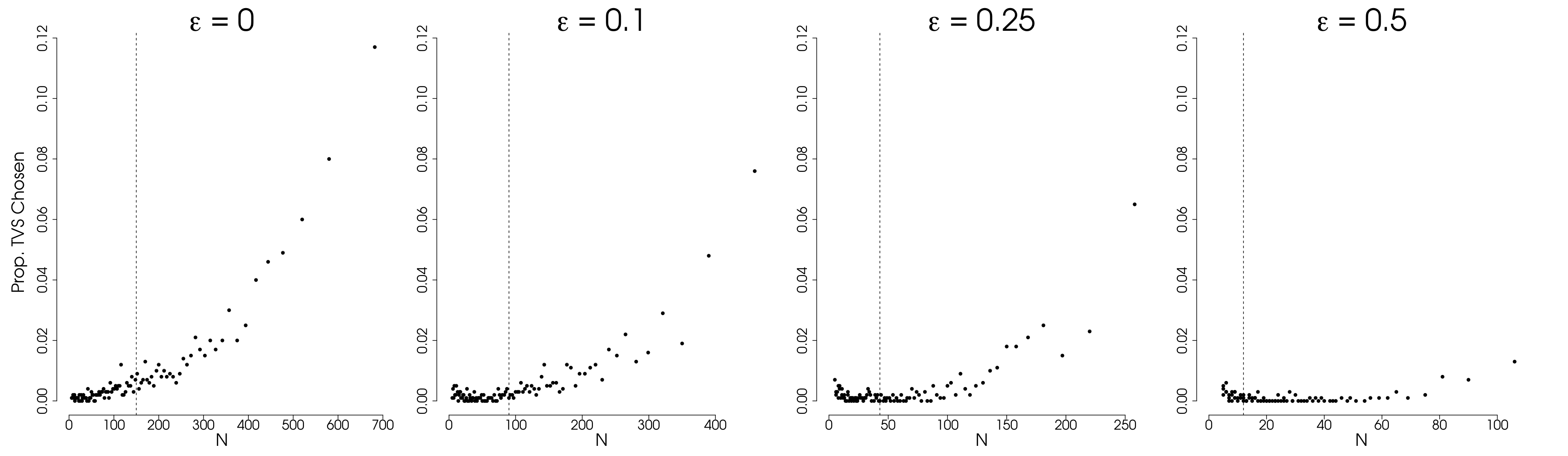}
\end{center}
\begin{flushleft}
\caption{
\textbf{Figure 4. Proportion of Trees Showing Support for Temporally-Varying Speciation Model.} Results from simulations showing the proportion of phylogenies for which the temporally-varying speciation (TVS) model (the `SPVAR' model of \cite{RaboskyLovette2008Evolution}) is preferred over a constant-rate birth-death (BD) model using AIC to select amongst the models. Only the TVS and BD models were compared. We used a $\Delta$AIC cutoff of 4 to favor a TVS model when the generating model was a constant-rate process. Extinction rate, $\epsilon=\mu/\lambda$, varies across the plots ($\epsilon=0,0.1,0.25,0.5$); speciation rate, $\lambda$, and total tree-depth, $\tau$ are held constant ($\lambda=1$ and $\tau=5$). All are plotted against the expected number of taxa across the cumulative distribution of probability densities (from 0.99 to 0.01). The dashed vertical line represents the expected value for $N$ under the simulating conditions. Each point represents 1000 simulations. (Results for $\epsilon =0.75$ and $\epsilon =1$ not shown.)} 
\end{flushleft}
\end{figure}

\begin{figure}[!ht]
\begin{center}
\includegraphics[width=6in]{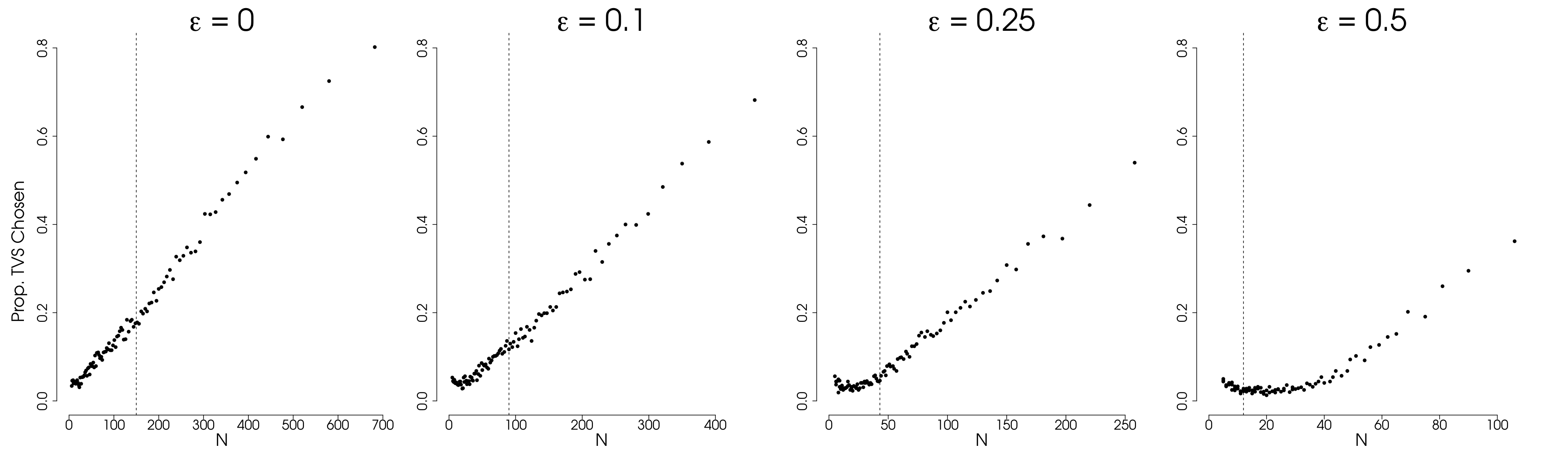}
\end{center}
\begin{flushleft}
\caption{
\textbf{Figure 5. Proportion of Trees Showing Support for Temporally-Varying Speciation Model Using Coalescent Approximation.} Results from simulations showing the proportion of phylogenies for which a temporally-varying speciation (TVS) model (Model 4a of \cite{Morlon2010}) is preferred over a constant-rate birth-death model (BD; Model 3 of \cite{Morlon2010}) using AIC to select amongst the models. Both the models were formulated according to a coalescent-based approximation of the likelihood \cite{Morlon2010}. We used a $\Delta$AIC cutoff of 4 to favor the TVS model when the generating model was a constant-rate process. Extinction rate, $\epsilon=\mu/\lambda$, varies across the plots ($\epsilon=0,0.1,0.25,0.5$); speciation rate, $\lambda$, and total tree-depth, $\tau$ are held constant ($\lambda=1$ and $\tau=5$). All are plotted against the expected number of taxa across the cumulative distribution of probability densities (from 0.99 to 0.01). The dashed vertical line represents the expected value for $N$ under the simulating conditions. Each point represents 1000 simulations. (Results for $\epsilon =0.75$ and $\epsilon =1$ not shown.)}
\end{flushleft}
\end{figure}

Stochastic bursts, which are more likely to have occured in `unusually large' trees, may be falsely inferred to have been caused by ecological processes (e.g. adaptive radiations). While the bursts are certainly `real', they are not attributable to any biological phenomena. If the background extinction is close to zero, then all methods investigated in this paper are susceptible to this false inference. However, when background extinction is relatively high, these stochastically generated slowdowns are not likely to be detected by the $\gamma$-statistic as the subsequent extinction removes the signal. As a summary statistic for the whole phylogeny, the power of the $\gamma$-statistic to detect slowdowns (stochastically or ecologically produced) when species turnover is present is very low \cite{Liow2010}, \cite{Quental2009}. It is fair to suggest that while there are some reasons to believe that significantly negative $\gamma$-values in the empirical literature may result from known statistical artifacts \cite{Revell2005},\cite{Cusimano2010},\cite{Fordyce2010}, the cause of their ubiquity remains unclear. We suggest that failing to sample cryptic species may contribute to this. Methods that do not deal directly with extinction and extinct lineages, like the diversity-dependent model \cite{RaboskyLovette2008PRSB} used here, show similar patterns of performance to the $\gamma$-statistic.

Our findings are also of relevance to studies examining the efficacy of diversification rate models and statistics. There are a multitude of ways to conduct simulations of birth-death models and Stadler \cite{Stadler2011Sys} has discussed the statistical properties of various conditioning schemes in detail. We caution theoreticians to pay close attention to these differences as conditioning simultaneously on the number of taxa and the age of the tree can produce phylogenies drawn from the `tails' of the distribution and thus prone to be bottom heavy. Such trees may lead to biased estimation of parameters and misconstrued inferences.

We suggest three future directions to address this issue of sampling bias. First, performing diversification analysis on megaphylogenies, including the clade of interest, may be advisable instead of examining clades in isolation. Investigating large clades while ignoring closely related groups that are less speciose may lead to spurious patterns of slowdowns in diversification. This will require further development of methods that relax the assumption of uniformity of the diversity dynamics across the tree (e.g. \cite{Alfaro2009}). We appreciate that for many groups, such an approach may not always be feasible so we urge researchers to at least be cautious in their interpretation of their results. Second, increased attention should be given to how lineages are defined; lineages that are currently denoted as subspecies may soon be considered ``good'' species \cite{Rosenblum2012}. Whether these lineages are included or not will influence the inference of slowdowns. Third, further investigation of the statistical properties of these models may allow researchers to be more confident that the patterns they observe represent truly meaningful variation.
% You may title this section "Methods" or "Models". 
% "Models" is not a valid title for PLoS ONE authors. However, PLoS ONE
% authors may use "Analysis" 

\section*{Methods}
All of our simulations were carried out using constant rates of speciation and extinction through time, so that any detection of a slowdown can be considered a Type-1 error. Tree simulations were conducted with Stadler's \cite{Stadler2011Sys} $\mathsf{TreeSim}$ $\mathsf{R}$ package, modified slightly for the purposes of computational efficiency. For each simulated phylogeny, we calculated the $\gamma$-statistic \cite{PybusHarvey2000} and fit five models of diversification: 1) a constant-rate birth-death model \cite{Nee1994}; 2) a diversity-dependent model \cite{RaboskyLovette2008PRSB}; 3) a time-varying speciation rate model \cite{RaboskyLovette2008Evolution}; 4) a constant-rate birth-death model using a coalescent approach \cite{Morlon2010}; and 5) a time-varying speciation rate model, also using a coalescent approach \cite{Morlon2010}. Models 1-3 were fit using the $\mathsf{R}$ package $\mathsf{laser}$ \cite{Rabosky2006}. Models 4 and 5 were fit using code from Morlon et al. \cite{Morlon2010}, also modified for computational efficiency.
%Note need to add Nee et al. 1994 to bibliography.

We simulated phylogenies conditioned on both the age of the tree $\tau$ and the number of taxa $N$ (for a discussion on various methods of conditioning simulated phylogenies, see \cite{Stadler2011Sys}). In order to simulate trees from different regions of the cumulative probability distribution (i.e. $1^{\mathrm{st}}$ percentile, $2^{\mathrm{nd}}$ percentile,$\ldots$, $99^{\mathrm{th}}$ percentile), we used the equations of Foote et al. \cite{Foote1999} for calculating $Pr(N | \lambda, \mu, \tau)$ where $\lambda$ is the instantaneous speciation rate per unit time and $\mu$ is the instantaneous extinction rate per unit time. The probability of observing a phylogenetic tree of age $\tau$, containing at least $n$ taxa, is given by the equations:

for $\lambda=\mu$

\begin{equation}\label{Foote1}
Pr(N \geq n\mid\lambda,\mu,\tau)=1-\frac{\mu(\exp[(\lambda-\mu)\tau] - 1) }{\lambda\exp[(\lambda-\mu)\tau]-\mu}-\sum_{i=1}^{n-1} (1-\alpha)(1-\beta)\beta^{i-1}
\end{equation}

where
\[\alpha=\frac{\mu(\exp[(\lambda-\mu)\tau] - 1) }{\lambda\exp[(\lambda-\mu)\tau]-\mu}\]

and
\[\beta=\frac{\lambda\alpha}{\mu}\]

and for $\lambda\neq\mu$

\begin{equation}\label{Foote2}
Pr(N \geq n\mid\lambda,\mu,\tau)=1-\frac{\lambda\tau}{(1+\lambda\tau)}-\sum_{i=1}^{n-1} \frac{(\lambda\tau)^{i-1}}{(1+\lambda\tau)^{i+1}}
\end{equation}

Note that in their formulation of these equations, Foote et al. \cite{Foote1999} use the paleontological convention of denoting $\lambda$ as $p$ and $\mu$ as $q$. We set $\lambda = 1$, $\tau = 5$ and varied $\mu$ ($\mu = \lbrace 0, 0.1, 0.25, 0.5, 0.75, 1.0 \rbrace$). Conditioning on the survival of at least one lineage to the present day, we used \eqref{Foote1} and \eqref{Foote2} to calculate the number of taxa associated with the percentiles of the cumulative distribution as discussed above. For each value of $\mu$ and each percentile ($p=0.01,0.02,\ldots,0 .99$), we simulated 1000 phylogenies under a constant-rate birth-death model (or in the case of $\mu = 0$, pure-birth). We also varied $\lambda$ and $\tau$ but the results of our analysis did not differ qualitatively. 

As stated above, the $\gamma$-statistic \cite{PybusHarvey2000} quantifies the temporal distribution of internode distances. We calculated $\gamma$ for each phylogeny using the $\mathsf{R}$ package $\mathsf{ape}$ \cite{Paradis2004}. Following the author's original recommendation \cite{PybusHarvey2000} and the conventions of the literature, we used a one-tailed test, such that $\gamma$ was considered to be significant if it was $< -1.645$. Again, as all phylogenies were simulated under a constant-rate process, we considered it to be a Type-1 Error if $\gamma$ was significantly negative.

The diversity-dependent (DD) model we used was the exponential declining model of Rabosky and Lovette \cite{RaboskyLovette2008PRSB} where speciation rate is modeled as a function of the number of contemporaneous lineages in the reconstructed phylogeny. If $N(t)$ the number of lineages at time $t$, $\lambda(0)$ is the initial (background) speciation rate and $\kappa$ describes the strength of diversity-dependence, $\lambda(t)$ can be described as the function 
\begin{equation}
\lambda(t)=\lambda(0)N(t)^{-\kappa}.
\end{equation}
In the DD model, background extinction rates are assumed to be 0. The decline in diversification rates at higher densities is thus only due to a decrease in speciation rate. 

To model time-varying speciation without explicitly considering diversity-dependence, we used the `SPVAR' model of Rabosky and Lovette \cite{RaboskyLovette2008Evolution} where speciation rate is an exponential function of time such that 
\begin{equation}
\lambda(t)=\lambda(0)e^{-xt}
\end{equation}
where $x$ is a constant describing the relationship. Here, extinction rate $\mu$ is explicitly modeled but assumed to be constant across the phylogeny.

As an alternative to using the full-likelihood equation for a time-varying speciation model {\cite{RaboskyLovette2008Evolution}, we also employed a recently derived approach \cite{Morlon2010} for fitting birth-death models, based on the coalescent process from population genetics \cite{GT1994}. Morlon et al. \cite{Morlon2010} model a population of species evolving under the Wright-Fisher process. Following Morlon et al. \cite{Morlon2010}, the likelihood $\mathcal{L}$ of observing the internode distances $g_{i}$ between node $i$ and node $i+1$ can be written as

\begin{equation}
\mathcal{L}(g_{i}) = \frac{i(i + 1)}{2} \frac{1}{N(u_{i})} \exp \left[ - \frac{i(i+1)}{2} \int_{u_{i} - g_{i}}^{u_{i}} \frac{1}{N(t)} dt \right]
\end{equation}
where $u_{i}$ is the distance between node $i$ and the present and $N(t)$ is the number of species (population size in population genetics) at time $t$ in the past. Note that this is an approximation of the likelihood as $N(t)$ is approximated by its expected value $\mathrm{E} \left[ N(t) \right]$ \cite{Morlon2010}; this was done to make the model analytically tractable. The constant-rate birth-death model we use corresponds to Model 3 from \cite{Morlon2010}. The time-varying speciation model was specified in the same way in the full-likelihood approach described above, such that $\lambda(t)=\lambda (0) e^{-xt}$. This corresponds to Model 4a from \cite{Morlon2010}.
	
For the model-based approaches, we compared a constant-rate birth-death model to each alternative model in our set. We used an AIC approach (sensu \cite{BA2002}) to select the best fit model between a constant-rate birth-death model and a time-varying speciation rate/diversity-dependent model. We emphasize for clarity that only 2 models were considered at a time. We used a $\Delta$AIC cutoff of 4 for the more parameter-rich model to be favored. While a $\Delta$AIC of 4 is an arbitrary value, we justify its use as it is conventionally used in phylogenetic comparative methods, following recomendations in \cite{BA2002}. All analyses were conducted in the $\mathsf{R}$ programming environment \cite{R}. The code modified from Stadler \cite{Stadler2011Sys} and Morlon et al. \cite{Morlon2010} will be available upon acceptance on M.W.P.'s personal website http://mwpennell.wordpress.com/

% Do NOT remove this, even if you are not including acknowledgments

\section*{Acknowledgments}
We thank Dan Wegmann, Graham Slater and Joseph Brown for providing some code to speed up the simulations and Jon Eastman, Tyler Hether, Jamie Voyles, Simon Uribe-Convers, Simon Ho and Lee Hsiang Liow for insightful comments on earlier versions of this manuscript. We also thank the RoHa lab and the Phylogenetics Reading Group (PuRGe) at the University of Idaho for stimulating discussion of these topics.

% The bibtex filename
\bibliography{burst_arxiv}

\end{document}